\documentclass[11pt]{article}
\usepackage[utf8]{inputenc}
\usepackage[english]{babel}
\usepackage{amsmath, amsthm, amssymb}
\usepackage{tikz}
\usepackage[margin=3.1cm,a4paper]{geometry}
\usepackage{bold-extra}

\usepackage{algorithm}
\usepackage{algcompatible}

\usepackage{thm-restate}

\newtheorem{theorem}{Theorem}
\newtheorem{corollary}[theorem]{Corollary}
\newtheorem{lemma}[theorem]{Lemma}

\theoremstyle{definition}
\newtheorem{definition}{Definition}

\theoremstyle{remark}
\newtheorem{claim}{Claim}

\usepackage{hyperref}

\hypersetup{
    colorlinks=true,
    linkcolor=blue,
    filecolor=black,
    citecolor=blue,
    urlcolor=black,
    pdftitle={Fast Euclidean Perfect Matching},
    }

\definecolor{mygreen}{rgb}{0, 0.7, 0}
\definecolor{myyellow}{rgb}{1, 0.8, 0}

\newcommand{\NN}{\textit{NN\,}}
\newcommand{\NRA}{\textsc{Node-Re\-duction-Algo\-rithm}}

\newcommand{\INRA}{\textsc{Itera\-ted-Node-Re\-duction-Algo\-rithm}}
\newcommand{\EFT}{\textsc{Edges-From-Tree}}
\newcommand{\EFNN}{\textsc{Edges-From-Nearest-Neighbor}}
\newcommand{\ECA}{\textsc{Even-Compo\-nent-Algo\-rithm}}
\newcommand{\MWPM}{\textit{MWPM\,}}
\newcommand{\R}{\mathbb{R}}
\newcommand{\PP}{\textsc{Place-Points}}

\title{Fast Approximation Algorithms for\\ Euclidean Minimum Weight Perfect Matching}

\author{Stefan Hougardy\thanks{Research Institute for Discrete Mathematics and Hausdorff Center for Mathematics, University of Bonn, Germany (hougardy@or.uni-bonn.de) funded by the Deutsche Forschungsgemeinschaft (DFG, German Research Foundation) under Germany's Excellence Strategy -- EXC-2047/1 -- 390685813 {https://orcid.org/0000-0001-8656-3418}}, Karolina Tammemaa\thanks{Hamburg University of Technology, Institute for Algorithms and 
Complexity, Hamburg, Germany (karolina.tammemaa@tuhh.de). Supported by Deutsche Forschungsgemeinschaft, DFG grant MN {59/4-1}}}

\begin{document}
\maketitle

\def\Rtwoexp{0.206}
\def\Rdexp{0.412}

\begin{abstract}
We study the Euclidean minimum weight perfect matching problem for $n$ points in the plane. 
It is known that any deterministic approximation algorithm 
whose approximation ratio depends only on $n$ requires at least $\Omega(n \log n)$ time.
We propose such an algorithm for the Euclidean minimum weight perfect matching problem with runtime $O(n\log n)$ and show that it has approximation ratio
$O(n^{\Rtwoexp})$. This improves the so far best known approximation ratio of $n/2$. 
We also develop an $O(n \log n)$ algorithm for the Euclidean minimum weight perfect matching problem in higher dimensions and show it has approximation ratio  $O(n^{\Rdexp})$ in all fixed dimensions.
\end{abstract}

\textbf{Keywords:} Euclidean matching · Approximation algorithms

\section{Introduction}

A \emph{perfect matching} in a graph $G$ is a subset of edges such that each vertex of $G$ is incident to exactly one edge in the subset.
When each edge $e$ of $G$ has a real weight $w_e$, then the \emph{minimum weight perfect matching problem} is to find a perfect matching $M$ that minimizes the weight $\sum_{e\in M}w_e$.
If the vertices are points in the Euclidean plane and we have a complete graph where each edge $e$ has weight $w_e$ equal to the Euclidean distance between its two endpoints, we call this the \emph{Euclidean minimum weight perfect matching problem}.

The Euclidean minimum weight perfect matching problem can be solved in polynomial time by applying Edmonds' blossom algorithm~\cite{edmonds1965maximum, edmonds1965paths} to the complete graph where the edge weights are the Euclidean 
distances between the edge endpoints. Gabow~\cite{Gab1972} and Lawler~\cite{lawler2001combinatorial}
have shown how Edmonds' algorithm can be implemented 
to achieve the runtime $O(n^3)$ on graphs with $n$ vertices. This implies an $O(n^3)$ runtime for the
Euclidean minimum weight perfect matching problem on point sets of size $n$.
By exploiting the geometry of the problem Vaidya~\cite{vaidya1988geometry} 
developed an algorithm for the Euclidean minimum weight perfect matching problem
with runtime $O(n^{\frac{5}{2}}\log n)$. In 1998, Varadarajan~\cite{varadarajan1998divide} 
improved on this result by presenting an $O(n^{\frac{3}{2}}\log^5(n))$ algorithm 
that uses geometric divide and conquer. 
This is the fastest exact algorithm currently known for the Euclidean minimum weight perfect matching problem.

Faster approximation algorithms for the Euclidean minimum weight perfect matching problem are known. In~\cite{vaidya1989approximate} Vaidya presented a $1+\varepsilon$-approximation algorithm with runtime $O(n^{\frac{3}{2}}\log^{\frac{5}{2}}n (1/\varepsilon^3) \sqrt{\alpha(n,n)})$, where $\alpha$ is the inverse Ackermann function and $\varepsilon \le 1$. Even faster approximation algorithms are known when  randomization is allowed. 
Arora~\cite{arora} presented a randomized $1+\varepsilon$ approximation algorithm with runtime $O(n\log^{O(1/\varepsilon)}n)$ and failure probability $1/2$. Varadarajan and Agarwal~\cite{Varadarajan_and_Agarwal} improved on this, presenting a randomized $1+\varepsilon$ approximation algorithm with runtime $O(n/\varepsilon^3 \log^6n)$. Rao and Smith~\cite{raoSmith} gave a constant factor $\frac{3}{2}e^{8\sqrt{2}}\approx 122\,905.81$ randomized approximation algorithm with runtime $O(n\log{n})$. In the same paper, they also propose a deterministic $\frac{n}{2}$-approximation algorithm with runtime $O(n\log n)$.

Grigoriadis and Kalantari~\cite{GK1986} have shown that any deterministic
approximation algorithm 
for the Euclidean minimum weight perfect matching problem with approximation factor depending only on $n$
needs to have a runtime of at least $\Omega(n\log{n})$. It is therefore natural 
to ask: \textit{What approximation ratio can be achieved in $O(n\log n)$ by a deterministic approximation algorithm for the Euclidean minimum weight perfect matching problem for $n$ points in the plane?}
In this paper we will significantly improve the approximation ratio of $n/2$  due to Rao and Smith~\cite{raoSmith}
and show:

\begin{restatable}{theorem}{approxtwodimensional}
For $n$ points in $\R^2$ there exists a deterministic $O(n^{\Rtwoexp})$-approximation algorithm for 
the Euclidean minimum weight perfect matching problem with runtime $O(n \log n)$.
\label{thm:approx2D}
\end{restatable}

Our algorithm is based on the idea of clustering the given points in the Euclidean plane 
into components of even cardinality. We show that there is a way to compute these components 
such that for all but one of the components we can find a perfect matching with 
small enough weight. We then apply our algorithm iteratively on this single remaining component. 
After sufficiently many iterations the number of points contained in the remaining 
component is small enough to apply an exact algorithm. 
Note that in the Euclidean minimum weight perfect matching problem we do not have the property
that the value of an optimum solution is at least as large as the optimum value for any even cardinality 
subset of the given points. 
Crucial for the analysis of our approach
is therefore that we can show that the size of the remaining component decreases faster than 
the value of an optimum solution for the remaining component increases.
The algorithm of Rao and Smith~\cite{raoSmith} can be extended to higher (fixed) dimension 
resulting in an approximation ratio of $n$ in runtime $O(n\log n)$.
Our approach also extends to higher (fixed) dimensions and we obtain the following result:

\begin{restatable}{theorem}{approxhigherdimension}
For any fixed dimension $d$ there exists a deterministic $O(n^{\Rdexp})$-approxi\-mation algorithm for 
the Euclidean minimum weight perfect matching problem in $\R^d$  with runtime $O(n \log n)$.
\label{thm:approxhigherdimension}
\end{restatable}

Theorem~\ref{thm:approx2D} and Theorem~\ref{thm:approxhigherdimension} improve earlier results we obtained in a preliminary version of this paper~\cite{HT2025}. There, we proved approximation factors of $O(n^{0.2995})$ and $O(n^{0.599})$ for the $2$-dimensional and the $d$-dimensional case. 
Crucial to obtain the improved results presented here is to allow the \NRA\ to make 
many iterations and that we are able to analyze this algorithm. 

The paper is organized as follows. In Section~\ref{sec: preliminaries} we introduce the nearest neighbor graph
and study its relation to minimum spanning trees and matchings. A basic ingredient to our main algorithm is 
a subroutine we call the \ECA. We explain this subroutine and prove some basic facts about it in 
Section~\ref{sec:ECA}. This subroutine was also used in the \emph{Even Forest Heuristic}
of Rao and Smith~\cite{raoSmith} that deterministically achieves in $O(n \log n)$ 
the so far best approximation ratio of $n/2$. 
We will apply the \ECA\ as a subroutine within our \NRA.
The idea of the \NRA\ is to partition a point set into some subsets of even cardinality and 
a single remaining set of points such that two properties hold. First, the even subsets should allow to compute
a short perfect matching within linear time. Second, the remaining set of points should be sufficiently small. 
In Section~\ref{Sec:node-reduction-algorithm} we will present our \NRA\ and will derive bounds for 
its runtime and the size of the remaining subset. The next idea is to iterate the \NRA\ on the remaining
subset until it becomes so small that we can apply an exact minimum weight perfect matching algorithm to it. 
We call the resulting algorithm the \INRA\ and analyze its runtime and approximation ratio in 
Section~\ref{sec:INRA} and also prove there Theorem~\ref{thm:approx2D}. 
In Section~\ref{sec:higherDimensions}
we extend the \INRA\ to higher (fixed) dimension and prove Theorem~\ref{thm:approxhigherdimension}.
Finally, in Section~\ref{sec:lowerbound} we provide a lower bound example for the \INRA.

\section{Preliminaries}
\label{sec: preliminaries}

A crucial ingredient to our algorithm is the so-called \emph{nearest neighbor graph}~\cite{PS1985}.
For a given point set in $R^d$ we first fix an arbitrary total ordering on all points, which we use to break ties when creating the nearest neighbor graph to avoid getting cycles.
We compute for each point all other points that have the smallest possible distance to this point. Among all these points we select, as its nearest neighbor, a point that is minimal with respect to the total ordering. 
Now we get the nearest neighbor graph by taking as vertices all points and adding an undirected edge between each point and its nearest neighbor without adding parallel edges. 
We will denote the nearest neighbor graph for a point set $V \subseteq \R^d$ by $\NN(V)$. Immediately from the definition we get that the nearest neighbor graph is a forest. It is well known that the nearest neighbor graph for a point set in $\R^d$ is a subgraph of a Euclidean minimum spanning tree for this point set~\cite{PS1985}. For $k\ge 2$ the nearest neighbor graph can be generalized to the $k$-nearest neighbor graph. To obtain this graph choose for each point its $k$ nearest neighbors (ties broken arbitrarily) and connect them with an edge.

Shamos and Hoey~\cite{SH1975} have shown that the nearest neighbor graph and a Euclidean minimum spanning tree 
for a point set of cardinality $n$ in $\R^2$ can be computed in $O(n \log n)$. 
For the nearest neighbor graph this result also holds in higher (fixed) dimension
as was shown by Vaidya~\cite{Vai1989}. The algorithm of Vaidya~\cite{Vai1989} 
even allows to compute the $k$-nearest neighbor graph for point sets in $\R^d$ in $O(n \log n)$
as long as $k$ and $d$ are fixed.

We call a connected component of a graph an \emph{odd connected component} if it has odd cardinality. 
Analogously, we define an \emph{even connected component}. We will denote by $\ell(e)$ the Euclidean length of an edge $e$ and for a set $E$ of edges we define $\ell(E) := \sum_{e\in E} \ell(e)$.
For a point set $V$ we denote a Euclidean minimum weight perfect matching for this point set by $\MWPM(V)$. 
Clearly, the point set $V$ must have even cardinality for a perfect matching to exist.
Throughout this paper by \emph{$\log $} we mean the logarithm with base~2.

There is a simple connection between the length of the nearest neighbor graph for a point set $V$ and 
a Euclidean minimum weight perfect matching for this point set:

\begin{lemma} For a point set $V\subseteq \R^d$ we have $\ell(\NN(V)) \le 2 \cdot \ell(\MWPM(V))$.
\label{lem:NNvsMWPM}    
\end{lemma}
\begin{proof}
In a Euclidean minimum weight perfect matching for a point set $V$ each point $v\in V$ is incident to an edge that is at least as long as the distance to a nearest neighbor of $v$. 
Assign to each point $v \in V$ the length of the edge that it is incident to in a Euclidean minimum weight perfect matching of $V$. This way a total length of $2 \cdot \ell(\MWPM(V))$ is assigned to 
the vertices. For each vertex the assigned edge length is at least as large as the distance to a 
nearest neighbor. Thus, by summing over all vertices we get $\ell(\NN(V)) \le 2 \cdot \ell(\MWPM(V))$.
\end{proof}

\section{The Even Component Algorithm}
\label{sec:ECA}

The currently best deterministic $O(n \log n)$ approximation algorithm for the Euclidean minimum weight
perfect matching problem is the Even Forest Heuristic due to Rao and Smith~\cite{raoSmith}. It achieves an approximation ratio of $n/2$ and tight examples achieving this approximation ratio are known~\cite{raoSmith}.
The Even Forest Heuristic first computes a minimum spanning tree of the graph. It then removes all 
so-called \emph{even edges}, where an edge is even if its removal splits the tree into two connected components of even cardinality. After removing all such edges, the remaining graph is a forest whose connected components all have even cardinality.
Within each connected component of the forest the Even Forest Heuristic then computes a Hamiltonian cycle
by first doubling all edges of the tree and then short-cutting a Eulerian cycle. 
One obtains a matching from the Hamiltonian cycle by taking every second edge. 
We also make use of this second part of the Even Forest Heuristic  
and call it the \ECA  \,(see Algorithm~\ref{alg:ECA}). 
In line~5 of this algorithm we shortcut the edges of a Eulerian cycle. 
By this we mean that we iteratively replace for three consecutive vertices $x,y,z$ the edges $xy$ and $yz$ by 
the edge $xz$ if the vertex $y$ has degree more than two.  
Notably, Rao and Smith applied this algorithm to a forest derived from a minimum spanning tree, we will apply it to a forest derived from the nearest neighbor graph.

\begin{algorithm}[ht]
\caption{~~\ECA}
\label{alg:ECA}
\begin{algorithmic}[1]
\Statex \textbf{Input: } a forest $F$ where all connected components have even cardinality
\Statex \textbf{Output: } a perfect matching $M$
\STATE  $M :=\emptyset$
\FOR{each connected component of $F$}
    \STATE double the edges in the component
    \STATE compute a Eulerian cycle in the component
    \STATE shortcut the edges in the Eulerian cycle to get a Hamiltonian cycle
    \STATE add the shorter of the two perfect matchings inside the Hamiltonian cycle to $M$.
\ENDFOR
\end{algorithmic}
\end{algorithm}

For completeness we briefly restate the following two results and their proofs from~\cite{raoSmith}:

\begin{lemma}[\cite{raoSmith}]
The \ECA\ applied to an even connected component 
returns a matching with length at most the total edge length of all edges in the even component.
\label{lem:evencomponentalgorithm}
\end{lemma}

\begin{proof}
By doubling all edges in a component we double the total edge length. 
The Eulerian cycle computed in line~4 has therefore exactly twice the length of the edges in the connected component.
The triangle inequality implies that shortcutting this cycle in line~5 cannot make it longer. The Hamiltonian cycle therefore
has length at most twice the length of all edges in the connected component. 
One of the two perfect matchings into which we can decompose the Hamiltonian cycle has length at most half of the length of the Hamiltonian cycle. Therefore, the length of the matching computed by the \ECA\ is upper bounded by the length of all edges in the connected component.
\end{proof}

\begin{lemma}[\cite{raoSmith}]
The \ECA\ has linear runtime. 
\label{lem:runtimeECA}
\end{lemma}
\begin{proof}
  We can use depth first search to compute in linear time the connected components and double the edges in all components.
  A Eulerian cycle can be computed in linear time using for example Hierholzer's algorithm~\cite{Hie1873}. 
  Shortcutting can easily be done in linear time by running along the Eulerian cycle.
  We find the smaller of the two matchings by selecting every second edge of the Hamiltonian cycle. 
\end{proof}

\section{The Node Reduction Algorithm}
\label{Sec:node-reduction-algorithm}

A central part of our algorithm is a subroutine we call the \emph{\NRA}. 
This algorithm gets as input a point set $V \subseteq \R^2$ of even cardinality and returns 
a subset $W\subseteq V$ and a perfect matching $M$ for $V\setminus W$. The idea of this algorithm is 
to first compute the nearest neighbor graph $\NN(V)$ of $V$. If $\NN(V)$ has many
odd connected components, then we will reduce this number by adding additional edges. 
More precisely, we will carry out up to $r$ rounds of iteratively adding edges to $\NN(V)$.
In round $i\in\{1, \ldots, r\}$ we will add edges if the current number of odd connected components is larger than $|V|/{x_i}$. 
We will later choose appropriate values for $r$ and $x_1, \ldots, x_r$. 
From each of the remaining odd connected components we remove one leaf vertex and put it into $W$. 
We are left with a set of even connected components and apply the \ECA\ to 
each of these. Algorithm~\ref{alg:Node-Reduction-Algorithm} shows the pseudo code of the
\NRA. It uses as a subroutine the algorithm 
\EFT\ which is shown in Algorithm~\ref{alg:Add-Edges-From-Tree-Algorithm}.

\begin{algorithm}[ht]
\caption{~~\NRA}
\label{alg:Node-Reduction-Algorithm}
\begin{algorithmic}[1]
\Statex \textbf{Input: } a set $V \subseteq \R^2$ of even cardinality, $r \in \mathbb{N}$, $x_1 < x_2 < \ldots < x_r$ with $x_i\in\mathbb{R}$
\Statex \textbf{Output: } $W \subseteq V$ and a perfect matching $M$ for $V\setminus W$
\STATE $G_0 := \NN (V)$                                            \label{line:NN-computation}
\STATE compute a Euclidean minimum spanning tree $T$ of $V$                \label{line:MST-computation}
\STATE $ i := 0$
\WHILE{$i < r$ and the number of odd connected components of $G_i$ is $> |V|/{x_{i+1}}$}  \label{line:while-loop}
     \STATE $i = i + 1$                                           \label{line:first-line-while-loop}
     \STATE $G_i := G_{i-1}\cup$ \EFT$(G_{i-1},T)$                 \label{line:last-line-while-loop}
\ENDWHILE
\STATE $W:=\emptyset$                                             \label{line:first-line-compute-W}
\FOR{each odd connected component of $G_i$}
    \STATE choose one leaf node in the component and add it to $W$
\ENDFOR                                                           \label{line:last-line-compute-W}   
\STATE $M := $ \ECA $(G_i[V\setminus W])$                           \label{line:matching-computation} 
\end{algorithmic}
\end{algorithm}

\begin{algorithm}[ht]
\caption{~~\EFT}
\label{alg:Add-Edges-From-Tree-Algorithm}
\begin{algorithmic}[1]
\Statex \textbf{Input: } a graph $G$ on vertex set $V\subseteq\mathbb{R}^2$ and a tree $T$ on $V$
\Statex \textbf{Output: } a subset of the edges of $T$
\STATE $S := \emptyset$
\STATE Choose a bijection $f:E(T)\to \{1,\ldots, |T|-1\}$ s.t.\ $f(e_1) < f(e_2)$ implies $l(e_1) \le l(e_2)$\label{line:function-f}
\FOR{each odd connected component of $G$}
     \STATE add to $S$ an edge from $T$ with minimum $f$-value that leaves this component
\ENDFOR
\STATE return the set $S$
\end{algorithmic}
\end{algorithm}

We start by giving a bound on the length of the edges returned by the subroutine \EFT.
\begin{lemma}
\label{lemma:subroutine-Edges-From-Tree}
    For a point set $V\subseteq \R^2$ let $G$ be a graph on $V$ and $T$ be a Euclidean minimum spanning tree on $V$.
    The subroutine \EFT$(G,T)$ returns a set of edges of length at most $2\cdot\ell(\MWPM(V))$.
\end{lemma}

\begin{proof}
    Let $S$ denote the edges returned from the subroutine  \EFT$(G,T)$.
    For each odd connected component of $G$ there must exist an edge $e$ in 
    $\MWPM(V)$ that leaves this component. By the cut property of minimum spanning trees the tree $T$ 
    contains an edge that leaves the component and has at most the length of $e$. As each edge in 
    $\MWPM(V)$ can connect at most two odd connected components we get the upper bound 
    $\ell(S) \le 2\cdot\ell(\MWPM(V))$. 
\end{proof}

In the following we analyze the performance and runtime of the \NRA.
One crucial part is to bound the size of the set $W$ returned by the \NRA.
For this it will be useful to assume a certain structure on the 
even and odd components in the graphs $G_i$ that are computed within the \NRA.
We make this more precise in the following definition.

\begin{definition}
Let $V\subseteq \mathbb{R}^2$ be a point set, $T$ a minimum spanning tree for $V$, and $r\in\mathbb{N}$. 
Set $G_0 := \NN(V)$ and recursively define $G_i := G_{i-1}\cup$ \EFT$(G_{i-1},T)$  for $i=1, \ldots, r$.
The set $V$ is \emph{$r$-well structured} if the following holds:
\begin{itemize}
    \item all connected components of $G_0$ have size $2$ or $3$
    \item for $i=1, \ldots, r$ each even connected component in $G_i$ is either an even connected component in $G_{i-1}$ or it consists of
    exactly two odd connected components of $G_{i-1}$
    \item for $i=1, \ldots, r$ each odd connected component in $G_i$ consists either of exactly one odd and one even connected component of $G_{i-1}$ or it consists of exactly three odd connected components of
    $G_{i-1}$
\end{itemize}
\end{definition}

\begin{figure}[ht]
\begin{tikzpicture}[scale=0.15] % first number was y, second x — swapped below

  \foreach \name/\x/\y in {
A/5/15,
B/10/24,
C/26/6,
D/38/39,
E/44/35,
F/45/30,
G/47/10,
H/48/45,
I/58/44,
J/65/41,
K/63/27,
L/66/23,
M/63/5,
N/75/21,
O/75/38,
P/81/32,
Q/81/3,
R/86/22,
S/92/39,
T/93/5,
U/95/25,
V/99/2
  } \node[circle, fill=black, inner sep=2pt] (\name) at (\x,\y) {};

 \draw[line width = 2, mygreen] (A)--(B) (C)--(G)--(M) (Q)--(T)--(V) (D)--(E)--(F) (H)--(I)--(J) (K)--(L)--(N) (O)--(P)--(S) (R)--(U);
  \draw[line width = 2, red] (E)--(H) (J)--(O) (N)--(R) (M)--(Q);
 \draw[line width = 2, myyellow] (P)--(R);
 \draw[thin, lightgray] (B)--(C) (L)--(M);
  \end{tikzpicture}
\vspace*{6mm}

\begin{tikzpicture}[scale=0.43]
  % baseline
  \draw (-0.5,0) -- (33.5,0);
  \foreach \x in {0,...,33}
    \draw (\x, -0.2)--(\x, 0.2);

  % 22 equidistant points named A..V
    \foreach \name/\x in {
A/0,
B/1,
C/5,
D/6,
E/7,
F/9,
G/10,
H/11,
I/15,
J/16,
K/17,
L/19,
M/20,
N/21,
O/23,
P/24,
Q/25,
R/28,
S/29,
T/30,
U/32,
V/33
  } \node[circle, fill=black, inner sep=2pt] (\name) at (\x,0) {};

 \draw[line width = 2, mygreen] (A)--(B) (C)--(D)--(E) (F)--(G)--(H) (I)--(J)--(K) (L)--(M)--(N) (O)--(P)--(Q) (R)--(S)--(T) (U)--(V);
 \draw[line width = 2, red] (E)--(F) (K)--(L) (N)--(O) (T)--(U);
 \draw[line width = 2, myyellow] (Q)--(R);
\end{tikzpicture}
\caption{(Top:) A $2$-well structured point set. The edges of the nearest neighbor graph $G_0$ are shown in green.
The graph $G_1$ contains in addition the red edges; the graph $G_2$ contains all colored edges. The two gray edges are edges of the minimum spanning tree. (Bottom:) The rearranged point set obtained after the first step of the construction described in the proof of Lemma~\ref{lemma:structure-of-V}.} 
\label{fig:well-structured-set}
\end{figure}

Figure~\ref{fig:well-structured-set} shows an example of a 2-well structured point set of size 22. The graph $G_0$ contains six connected components of size~3 and two connected components of size~2. 
The graph $G_1$ is obtained from $G_0$ by adding the four red edges. $G_1$ contains two even connected components, one of size~2 and one of size~6. The connected component of size~2 was already a connected component in $G_0$. The connected component of size~6 in $G_1$ is obtained by joining two odd connected components of $G_0$. Moreover, $G_1$ contains two odd connected components, one of size 5 and one of size 9.
The connected component of size~5 in $G_1$ consists of one odd and one even connected component from $G_0$; the connected component of size~9 in $G_1$ consists of three odd connected components of $G_0$. 
Finally, the graph $G_2$ is obtained from $G_1$ by adding the yellow edge between the two 
odd connected components of $G_1$. Note that all connected components of $G_2$ are even and therefore the point set is not only 2-well-structured but it is $r$-well structured for all $r\in\mathbb{N}$.

The following lemma shows that for each point set $V$ there exists an $r$-well structured point set $V'$ with the 
same size that results in the same number of odd connected components in each iteration of the \NRA.
This result will allow us to prove in Corollary~\ref{cor:well-structured} that for analyzing
the \NRA\ it is enough to consider $r$-well structured point sets.  

\begin{lemma}
\label{lemma:structure-of-V}
Let $V\subseteq \R^2$ be a point set,  $r\in \mathbb{N}$ with $r\ge 0$, and  
$2 < x_1 < x_2 < \ldots < x_r$ with $x_i\in\mathbb{R}$ such that the \NRA\ makes exactly $r$ iterations. 
Let $G_i$ be the graph computed by the \NRA\ in iteration $i$ of the while-loop. 
Then there exists a well structured point set $V'\subseteq \mathbb{R}^2$ with $|V|=|V'|$ 
such that for each $i=0, \ldots, r$ the graph $G_i'$ computed by the \NRA\ on input $V'$ has the same 
number of odd connected components as the graph $G_i$. 
\end{lemma}

\begin{proof} 
We will prove this result in two steps. First we show that one can rearrange 
the points in $V$ so that they all lie on a horizontal line and the \NRA\ will still have the same set of vertices in each connected component of the graphs $G_i$ for $i=0, \ldots, r$. This is clear for the vertices of a single connected component of the graph $G_0$: if the cardinality of the component is $k$ then for some number $a\in\mathbb{N}$ we can place the $k$ vertices at coordinates $(a+1, 0), (a+2,0), \ldots, (a+k, 0)$ and just have to make sure that all other vertices of $V$ are placed left of $(a-1, 0)$ or right of $(a+k+2, 0)$.  By defining an appropriate tie breaking function the nearest neighbor graph $G_0$ will contain exactly all edges of length~1 and therefore the $k$ vertices still lie in a connected component of $G_0$. Using a recursive approach this idea can easily be extended to work for all connected components of all graphs $G_i$ for $i=0, \ldots, r$.

For a more formal description we show how to rearrange the points in $V$ such that they all 
lie on a horizontal line and for $i=0, \ldots, r$ each connected component of the graph $G_i$ computed by the \NRA\ on the rearranged point set contains the same vertices as the connected components of the graph $G_i$
for the original point set $V$. For this we use the algorithm \PP~(Algorithm~\ref{alg:placepoints}). 
It gets as input an integer $a$, a point set $S$ and an index $i$. 
Starting at $x$-coordinate $a$ the algorithm \PP\ recursively places all 
points in $S$ on a horizontal line according to the connected components in the graph $G_j$ for $j\le i$.
All points in $S$ that belong to the same connected component of $G_i$ are placed recursively
by \PP\ such that each point has distance $1$ to some point from the same connected component but distance of at least $i+2$ to points from a different connected component of $G_i$. 
At the lowest level of the recursion the points in a connected component of $G_0$ are placed one after the other on a horizontal line such that each point has distance one to the previous one. The first point of the next connected component of $G_0$ is then placed with larger distance.

\begin{algorithm}[ht]
\caption{~~\PP}
\label{alg:placepoints}
\begin{algorithmic}[1]
\Statex \textbf{Input: } an integer $a$, a set of points $S$, and an index $i$
\Statex \textbf{Output: } $x$-coordinates for all points in $S$
%\STATE  $M :=\emptyset$
\FOR{each connected component $C$ of $G_i$ within $S$}
 \IF{$i=0$}
    \STATE place the points in $C$ at $x$-values $a, a+1, \ldots a+|S|-1$
    \ELSE
      \STATE call \PP($a$, $V(C)$, $i-1$) 
  \ENDIF
\STATE $a:= i+2~+$ the largest $x$-value of a point in $V(C)$
\ENDFOR
\end{algorithmic}
\end{algorithm}

We get all the coordinates of the rearranged points by the call \PP$(0, V, r)$. 
As a tie breaking function for the nearest neighbor graph, we use an ordering of the vertices on the line from left to right.
A minimum spanning tree on the rearranged point set is simply obtained by taking all edges between neighboring points on the line. 
As a tie breaking function $f$ for edges of equal length, needed in line~\ref{line:function-f} of the algorithm \EFT, we 
use the ordering of the edges of the minimum spanning tree from left to right. 
Then, the \NRA\ applied to the rearranged point set will have exactly the same vertices in each connected component of $G_i$ as for the original point set. 
Figure~\ref{fig:well-structured-set} shows a possible rearrangement obtained by the above procedure 
for the set of 22 points shown in the figure.  

Let us call $V''$ the set of rearranged vertices we obtained from $V$ by the above procedure.  
In a second step we modify the point set $V''$ to obtain the $r$-well structured point set $V'$ 
by relocating some of the points. We start by setting $V' := V''$ and use the following recursive procedure for $i=r, \ldots, 0$
to modify the location of some points in $V'$ (see Figure~\ref{fig:compressed-set} for an example). To avoid a special treatment for the case $i=0$ we assume that $G_{-1}$ is the graph where each vertex is an odd component. 
If a connected component $C$ of $G_i$ contains more than three odd connected components of $G_{i-1}$ then we choose the two left most odd connected components
$C_1$ and $C_2$ of $G_{i-1}$  that lie in $C$ and translate the points in $C_1$ and $C_2$ to the right of all points in $V'$
such that the points in $C_1$ have distance $r+2$ to the right most point in $V'$ and the points in $C_2$ are right of all the points in $C_1$ and have distance $i+1$ to the right most point in $C_1$ (see Figure~\ref{fig:compressed-set}~b)).
Then we compact the space that the points in $C_1$ and $C_2$ had occupied before on the line. 
If the left most point of $C_1$ was at position $a$ and the right neighbor of the right most point in $C_1$ is at position 
$b$ then we shift all points that lie right of $a$ by $b-a$ to the left. We do the same for $C_2$ (see Figure~\ref{fig:compressed-set}~c)).
We apply this procedure as long as possible to $G_i$. 
We apply as long as possible a similar procedure to all odd connected components of $G_i$ that contain an even connected component of $G_{i-1}$ and more than one odd connected component of $G_{i-1}$. 
We apply a similar procedure to all even components of $G_i$ that contain more than 2 odd connected component of $G_{i-1}$ and no even connected component of $G_{i-1}$. 
We also apply a similar procedure to all even connected components of $G_i$ that contain at least 2 odd connected component of $G_{i-1}$ and one even connected component of $G_{i-1}$.
Figure \ref{fig:compressed-set} shows how this method would rearrange a given point set of size 15.

\begin{figure}[ht]

\def\drawline#1{\node[label=left: #1)] (dummy) at ( -0.7,0) {};
                \draw (-0.5,0) -- (30.5,0);
                \foreach \x in {0,...,30}
                  \draw (\x, -0.17)--(\x, 0.17);
}

\begin{tikzpicture}[scale=0.450]
  \drawline{a}
  
  \foreach \name/\x in {A/0,B/1,C/4,D/5,E/7,F/8,G/9,H/10,I/11,J/13,K/14,L/15,M/17,N/18,O/19} 
    \node[circle, fill=black, line width = 0, inner sep=1.5pt] (\name) at (\x,0) {};

  \draw[line width = 2, mygreen] (A)--(B) (C)--(D) (E)--(F)--(G)--(H)--(I) (J)--(K)--(L) (M)--(N)--(O);
  \draw[line width = 2, red] (D)--(E) (I)--(J) (L)--(M);
\end{tikzpicture}

\begin{tikzpicture}[scale=0.450]
  \drawline{b}
  
  \foreach \name/\x in {A/0,B/1,C/4,D/5,E/22,F/23,G/24,H/25,I/26,J/28,K/29,L/30,M/17,N/18,O/19} 
    \node[circle, fill=black, line width = 0, inner sep=1.5pt] (\name) at (\x,0) {};

  \draw[line width = 2, mygreen] (A)--(B) (C)--(D) (E)--(F)--(G)--(H)--(I) (J)--(K)--(L) (M)--(N)--(O);
\end{tikzpicture}

\begin{tikzpicture}[scale=0.450]
  \drawline{c}

  \foreach \name/\x in {A/0,B/1,C/4,D/5,E/12,F/13,G/14,H/15,I/16,J/18,K/19,L/20,M/7,N/8,O/9} 
    \node[circle, fill=black, line width = 0, inner sep=1.5pt] (\name) at (\x,0) {};

  \draw[line width = 2, mygreen] (A)--(B) (C)--(D) (E)--(F)--(G)--(H)--(I) (J)--(K)--(L) (M)--(N)--(O);
  \draw[line width = 2, red] (D)--(M) (I)--(J);
\end{tikzpicture}

\begin{tikzpicture}[scale=0.450]
  \drawline{d}
  
  \foreach \name/\x in {A/0,B/1,C/4,D/5,E/23,F/24,G/14,H/15,I/16,J/18,K/19,L/20,M/7,N/8,O/9} 
    \node[circle, fill=black, line width = 0, inner sep=1.5pt] (\name) at (\x,0) {};

  \draw[line width = 2, mygreen] (A)--(B) (C)--(D) (E)--(F) (G)--(H)--(I) (J)--(K)--(L) (M)--(N)--(O);
\end{tikzpicture}

\begin{tikzpicture}[scale=0.450]
  \drawline{e}

  \foreach \name/\x in {A/0,B/1,C/4,D/5,E/21,F/22,G/12,H/13,I/14,J/16,K/17,L/18,M/7,N/8,O/9} 
    \node[circle, fill=black, line width = 0, inner sep=1.5pt] (\name) at (\x,0) {};

  \draw[line width = 2, mygreen] (A)--(B) (C)--(D) (E)--(F) (G)--(H)--(I) (J)--(K)--(L) (M)--(N)--(O);
  \draw[line width = 2, red] (D)--(M) (I)--(J);
\end{tikzpicture}
\caption{
Relocation of a point set to create a 1-well structured set. 
a) shows the start state with the graph $G_1$ containing the red and green edges while
$G_0$ contains the green edges only. In b) two odd components of $G_0$ have been 
relocated to the right. In c) compaction has taken place. In d) two points from an odd
connected component of size~5 in $G_0$ have been relocated to the right. e) shows the final state after compaction. 
%The gray edges are edges of the minimum spanning tree.
}
\label{fig:compressed-set}
\end{figure}

We start applying these procedures to all connected components of $G_r$ and then continue with
the connected components of $G_{r-1}, G_{r-2}, \ldots, G_0$. 
When we apply these procedures to all connected components of $G_i$ we do not change the relative positions of points in the
same connected component of $G_{i-1}$.  Therefore, for each $i=0, \ldots, r$ the \NRA\ applied to $V'$ will have exactly the same number of odd connected components in $G_i'$ as the graph $G_i$.

\end{proof}

\begin{corollary}
\label{cor:well-structured}
For proving a bound on the size of the set $W$ returned  after $r$ iterations by the \NRA\ it is enough to prove a bound for $r$-well structured point sets.     
\end{corollary}

\begin{proof}
    The size of the set $W$ returned by the \NRA\ after $r$ iterations is equal to the number of odd connected components in the graph $G_r$. By Lemma~\ref{lemma:structure-of-V} there exists an $r$-well structured 
    point set $V'$ with $|V'|= |V|$ such that the set $W'$ returned by the \NRA\ on input set $V'$ has 
    the same size as the set $W$. 
    Thus, a bound on the size of $W'$ also yields the same bound on the size of $W$.
\end{proof}

The next lemma gives a bound on the length of the matching $M$ and the size of the set $W$ returned by the
\NRA\ depending on the number of calls of the subroutine \EFT.

\begin{lemma}
\label{lemma:bounds-for-remaining-vertices}
For a point set $V\subseteq \R^2$,  $r\in \mathbb{N}$ with $r\ge 0$, $1 < x_1 < x_2 < \ldots < x_r$ with $x_i\in\mathbb{R}$, let $W$ be the point set and $M$ be the matching returned by the \NRA. 
If the subroutine \EFT\ is called $q$ times within the \NRA\ then we have:
\label{lem:NRA}
\begin{enumerate}
\item  [(a)] $|W|\le $ 
$\begin{cases}
    |V|/x_{q+1} & \text{if } q<r\\
    |V|/ \frac{3}{1 - 2 \sum_{i=1}^{r} \frac{1}{x_i}} & \text{if } q =r
\end{cases}$
\item  [(b)] $\ell(M)\le (2 q+2) \cdot\ell(\MWPM(V))$
\item  [(c)] $\ell(\MWPM(W))\le (2 q+3)\cdot\ell(\MWPM(V))$
\end{enumerate}
\end{lemma}

\begin{proof}
We first prove statement~$(a)$ for $q<r$. In this case the while-loop in the 
\NRA\ is terminated because $G_q$ has at most $|V|/x_{q+1}$ odd connected components.
By lines~\ref{line:first-line-compute-W}--\ref{line:last-line-compute-W} the set $W$ contains exactly one node from each odd connected component of $G_q$. We therefore get $|W|\le |V|/x_{q+1}$. \medskip

We now prove statement~$(a)$ for the case $q =r$. By Corollary~\ref{cor:well-structured} we may 
assume that the point set $V$ is $r$-well structured. The graph $G_i$ contains the graph $G_{i-1}$ as a subgraph.
Moreover, the graph $G_0$ is $\NN(V)$. Therefore, for all $0\le i\le r$ each even connected component in $G_i$ contains at least two vertices from $V$ and each odd connected component in $G_i$ contains at least three vertices from $V$.

In the following we denote by $e_i$ the number of even connected components in $G_i$. Similarly, $o_i$ denotes the number of odd connected components in $G_i$. 

\begin{claim}
    For $1\le i \le r$ we have: $2o_{i-1} + 3 o_i + 2 e_i \le |V|$.
    \label{claim:V-bound}
\end{claim}

We prove this claim by showing that within $V$ we can choose three pairwise disjoint sets of size 
$2o_{i-1}$, $3 o_i$, and $2 e_i$. Within each odd connected component of $G_{i-1}$ we choose two vertices. 
This results in a set of size $2o_{i-1}$. If an even connected component in $G_i$ contains an even connected component from $G_{i-1}$ we can choose two vertices from this even connected component of $G_{i-1}$. Otherwise,
the even connected component in $G_i$ must contain at least two odd connected components from $G_{i-1}$ and we can choose from two such odd connected components a vertex that is distinct to the two already chosen vertices.
This gives us in total a set of $2 e_i$ vertices. 
If an odd connected component in $G_i$ contains at least three odd connected components from $G_{i-1}$, then we can choose from three of these odd connected components a vertex that is different to the two already chosen vertices.
Otherwise an odd connected component
in $G_i$ contains exactly one odd connected component from $G_{i-1}$ and one even connected component from $G_{i-1}$. We can now choose one vertex from the odd connected component and two vertices from the even connected component. Thus, for each odd connected component in $G_i$ we can choose three vertices that are different from vertices we have chosen before. Therefore we get $3 o_i$ additional vertices. This proves the claim.\medskip

For $i=1, \ldots, r$ we partition the odd and even connected components of $G_i$ each into two sets as follows:
\begin{itemize}
\item $e_{i,0}$ denotes the number of even connected components in $G_i$ containing no odd connected components of $G_{i-1}$.
\item $e_{i,2}$ denotes the number of even connected components in $G_i$ containing exactly two odd connected components of $G_{i-1}$.
\item $o_{i,1}$ denotes the number of odd connected components in $G_i$ containing exactly one even and one odd 
connected component of $G_{i-1}$.
\item $o_{i,3}$ denotes the number of odd connected components in $G_i$ containing exactly three odd 
connected components of $G_{i-1}$.  
\end{itemize}

Immediately from this definition and because we have an $r$-well structured point set we get $e_i = e_{i,0} + e_{i,2}$ and $o_i = o_{i,1} + o_{i,3}$. 
Moreover, we have 
$$ o_{i,1} + 2 e_{i,2} + 3 o_{i,3} = o_{i-1} \text{~~~and~~~} e_{i,0} + o_{i,1} = e_{i-1}.$$
Using these two equations we get
\begin{equation}
e_i= e_{i,0} + e_{i,2} = e_{i-1} - o_{i,1} + o_{i-1} - o_{i,1} - e_{i,2} - 3 o_{i,3} = e_{i-1} + o_{i-1} - 2 o_i - e_{i,2} - o_{i,3}
\label{eq:ei}
\end{equation}

\begin{claim}
    For $1\le i \le r$ we have $o_{i-1} - o_i = 2(e_{i,2}+o_{i,3})$
\label{claim:diff-odds}
\end{claim}
We know that $o_{i,1}+2e_{i,2}+3o_{i,3}=o_{i-1}$ and $o_i=o_{i,1}+o_{i,3}$ giving us $o_{i-1} - o_i = 2(e_{i,2}+o_{i,3})$. This proves the claim.\medskip

\begin{claim}
    For $1\le i \le r$ we have $o_i + 2e_i = -2o_i + (o_{i-1} + 2e_{i-1})$
    \label{claim:recursive-statement}
\end{claim}
By equation (\ref{eq:ei}) and Claim~\ref{claim:diff-odds} we have
\begin{eqnarray*}
 o_i + 2e_i  & = &  o_i +  2(e_{i-1}  + o_{i-1} - 2o_i - e_{i,2} - o_{i,3}) \\
             & = & -3o_i + 2e_{i-1} + 2o_{i-1} - 2(e_{i,2} + o_{i,3})  \\
             & = & -3o_i + 2e_{i-1} + 2o_{i-1} - (o_{i-1} - o_i)  \\
             & = & -2o_i + (o_{i-1} + 2e_{i-1})   
\end{eqnarray*}

\begin{claim}
For $r\in \mathbb{N}$ we have $3o_r \le  |V|  -2 \displaystyle\sum_{t=0}^{r-1}o_{t}$.
\label{claim:odd-bound}
\end{claim}
For $r=1$ we get from Claim~\ref{claim:V-bound} that:
\begin{eqnarray*}
3o_1 \le  |V|-2 o_{0} -2 e_{1} \le |V|-2 o_{0}
\end{eqnarray*}

For $r\ge 2$ we get by recursively applying Claim~\ref{claim:recursive-statement} and Claim~\ref{claim:V-bound} that:
\begin{eqnarray*}
3o_r &\le &  o_{r-1} + 2 e_{r-1} \\
     &=&  -2o_{r-1} + (o_{r-2}  +2 e_{r-2})\\
     &=&  -2\sum_{t=1}^{r-2}o_{r-t} + o_{1}  + 2 e_{1}\\
     &\le&  -2\sum_{t=1}^{r-2}o_{r-t} + o_{1} + |V| -2o_{0} -3 o_1 \\
     &=&|V|  -2 \sum_{t=0}^{r-1}o_{t}     
\end{eqnarray*}
This proves the claim.\medskip

As the algorithm did not terminate in earlier iterations we have for $1\le i\le r$ $o_{i-1} > |V|/x_i$. Thus, using Claim \ref{claim:odd-bound} we get:

\begin{eqnarray*}
o_r &\le&  \frac13 (|V| -2 \sum_{i=0}^{r-1}o_{i} ) \\
    &\le&  \frac13 (|V| -2 |V| \sum_{i=0}^{r-1}1/x_{i+1} ) \\
    & = & |V| \frac{ 1- 2 \sum_{i=1}^{r}1/x_{i}}{3}\\
     & = &  |V| / \frac{3}{1- 2 \sum_{i=1}^{r}1/x_{i}} 
\end{eqnarray*}

This finishes the proof of statement~$(a)$.\medskip

We now prove statement $(b)$.
Let $S$ be the set of all edges added to the graph $G$ 
in the $q$ iterations of the while-loop. 
From Lemma~\ref{lemma:subroutine-Edges-From-Tree} we know that the total length $\ell(S)$ of all edges in $S$ 
is at most $2 q \cdot\ell(\MWPM(V))$. In line~\ref{line:matching-computation} of the \NRA , the matching $M$ is computed 
by the \ECA\ on a subset of the edges in $\NN(V) \cup S$.
Lemma~\ref{lem:evencomponentalgorithm} therefore implies $\ell(M) \le \ell(\NN(V)) + \ell(S)$.
Now Lemma~\ref{lem:NNvsMWPM} together with the above bound on $l(S)$ yields
\begin{eqnarray*}
\ell(M) &\le& \ell(\NN(V)) + \ell(S) \\
        &\le& 2\cdot\ell(\MWPM(V)) + 2q\cdot\ell(\MWPM(V)) ~=~ 2(q+1)\cdot\ell(\MWPM(V)).
\end{eqnarray*}

Finally, we prove statement~$(c)$.
Let $H$ be the graph obtained from $\NN(V)$ by adding the edges of $S$ and $\MWPM(V)$. 
All connected components in $H$ have even cardinality as by definition $H$ contains a perfect matching.
This implies that each connected component of $H$ contains an even number of odd connected components
from $\NN(V) \cup S$. As $W$ contains exactly one point from each odd connected component of
$\NN(V)\cup S$, this implies that each connected component of $H$ contains an even number of vertices of $W$.
Within each connected component of $H$ we can therefore pair all vertices from $W$ and connect each pair 
by a path. By taking the symmetric difference of all these paths we get a set of edge disjoint paths in $H$
such that each vertex of $W$ is an endpoint of exactly one such path. The triangle inequality implies 
that an edge connecting the two endpoints of such a path is at most as long as the path. 
If we replace each path by the edge connecting its two endpoints we get a perfect matching for $W$. 
Therefore, we know that the total length of all these edge disjoint paths in $H$ is an upper
bound for $\ell(\MWPM(W))$. As the total length of all these edge disjoint paths is
bounded by the total length of all edges in $H$, we get by using Lemma~\ref{lem:NNvsMWPM}:
\begin{eqnarray*}
\ell(\MWPM(W)) &\le& \ell(E(H)) \\
               &\le& \ell(NN(V)) + \ell(S) + \ell(MWPM(V)) \\
               &\le& (2q+3) \cdot\ell(\MWPM(V)).
\end{eqnarray*}
(Part of this argument would easily follow from the theory of $T$-joins which we avoid to introduce here.) 
\end{proof}

\begin{lemma}
    For constant $r$ the \NRA\ (Algorithm~\ref{alg:Node-Reduction-Algorithm}) on input $V$ with $|V| = n$ has runtime $O(n\log n)$.
    \label{runtime:NRA}
\end{lemma}
\begin{proof}
In line~\ref{line:NN-computation} of the \NRA\ the nearest neighbor graph can be computed in $O(n\log n)$ as 
was proved by Shamos and Hoey~\cite{SH1975}. As the nearest neighbor graph has a linear number of edges
we can use depth first search to compute its connected components and their parity in $O(n)$.
The minimum spanning tree $T$ for $V$ in line~\ref{line:MST-computation} of the algorithm  can be computed in $O(n\log n)$
by using the algorithm of Shamos and Hoey~\cite{SH1975}.
To compute the set $S$ in the subroutine \EFT , simply run through all edges of $T$ 
and store for each odd connected component the shortest edge leaving that component. 
We have $r$ calls to the subroutine \EFT. As $T$ has $n-1$ edges we
get a total bound of $O(n)$ for all calls to the subroutine \EFT.
Choosing a leaf node in a connected component which is a tree can be done in time proportional to the
size of the connected component; thus the total runtime of lines~\ref{line:first-line-compute-W}--
\ref{line:last-line-compute-W} is $O(n)$. 
Finally, by Lemma~\ref{lem:runtimeECA} the runtime of the \ECA\ is linear in the size of
the input graph. Therefore, line~\ref{line:matching-computation} requires $O(n)$ runtime. 
Summing up all these time complexities gives us a time complexity $O(n\log{n})$.
\end{proof}

\section{Iterating the \NRA}
\label{sec:INRA}

The \NRA\ (Algorithm~\ref{alg:Node-Reduction-Algorithm}) on input $V\subseteq \R^2$ returns a set \mbox{$W\subseteq V$} and a perfect matching on $V\setminus W$.
The idea now is to iterate the \NRA\ on the set $W$ of unmatched vertices. 
By Lemma~\ref{lem:NRA}$(a)$ we know that after each iteration the set $W$ shrinks by at least a 
constant factor. 
Therefore, after a logarithmic number of iterations the set $W$ will be empty. 
However, we can do a bit better by stopping as soon as the set $W$ is small enough to 
compute a Euclidean minimum weight perfect matching on $W$ in $O(n \log n)$ time.
We call the resulting algorithm the \INRA, see Algorithm~\ref{alg:Iterated-Node-Reduction-Algorithm}.
In line~3 of this algorithm we apply the \NRA\ to the point set $V_j$.
This gives us a matching we denote by $M_j$ and a set of unmatched points which we denote by $V_{j+1}$.

\begin{algorithm}[ht]
\caption{~~\INRA}
\label{alg:Iterated-Node-Reduction-Algorithm}
\begin{algorithmic}[1]
\Statex \textbf{Input: } a set $V \subseteq \R^2$ of even cardinality, $\varepsilon > 0$,  $r \in \mathbb{N}$, $x_1 < x_2 < \ldots < x_r$ with $x_i\in\mathbb{R}$
\Statex \textbf{Output: } a perfect matching $M$ for $V$
\STATE $V_1 := V$, $j:=1$
\WHILE {$|V_j| > |V|^{2/3-\varepsilon}$}                                    \label{line:INRA:begin-while}
\State $V_{j+1}, M_{j}$ $\gets$  \NRA$(V_j,  r, x_1, x_2, \ldots, x_r)$  \label{line:INRA-calls-NRA}
\State $j := j + 1$
\ENDWHILE                                                                \label{line:INRA:end-while}
\State $M := \MWPM(V_{j}) \cup M_1 \cup M_2 \cup \ldots \cup M_{j-1}$                \label{line:INRA:matching-computation}
\end{algorithmic}
\end{algorithm}

Clearly, the \INRA\ returns a perfect matching on the input set $V$.
The next lemma states the runtime of the \INRA.

\begin{lemma}
\label{runtime-INRA}
    The \INRA\ (Algorithm~\ref{alg:Iterated-Node-Reduction-Algorithm}) on input $V$ with $|V| = n$, fixed $r\in \mathbb{N}$ and $x_1 > 2$ has runtime $O(n\log n)$.
\end{lemma}
\begin{proof}
    We have $|V_1| = |V|$ and because of $x_1 > 2$, Lemma~\ref{lem:NRA}(a) implies $|V_j|  < |V|/2^{j-1}$
    for $j \ge 2$. 
    By Lemma~\ref{runtime:NRA} the runtime of the \NRA\ on the set $V_j$ is $O(|V_j|\log |V_j|)$.
    The total runtime for lines~2--5 of the \INRA\ is therefore bounded by
    $$O\left(\sum_{j=1}^\infty |V_j| \log |V_j|\right) = 
      O\left(\log |V| \cdot \sum_{j=1}^\infty \frac{|V|}{2^{j-1}}\right) = O(|V| \log |V|).$$
    In line~6 of the \INRA\ we can use the algorithm of 
    Varadarajan~\cite{varadarajan1998divide} to compute a Euclidean minimum weight perfect matching on $V_j$.
    For a point set of size $s$ Varadarajan's algorithm has runtime 
    $O(s^{\frac{3}{2}}\log^5(s))$. As the set $V_j$ in line~6 of the \INRA\ has size
    at most $|V|^{2/3-\varepsilon}$ we get a runtime of 
    $$O\left((|V|^{2/3-\varepsilon})^{\frac{3}{2}}\log^5(|V|^{2/3-\varepsilon})\right) =
    O\left(\frac{|V|}{|V|^{\frac{3 \epsilon}{2}}}\log^5(|V|^{2/3-\varepsilon})\right)
    = O\left(|V|\right).$$
    In total the \INRA\  has runtime $O(n \log n)$.
\end{proof}

We now analyze the approximation ratio of the \INRA. 
For this we will have to choose appropriate values for $x_1, \ldots, x_r$. Our analysis 
will yield the best approximation ratio if we choose the $x_i$ in such a way that 
\begin{equation}\label{condition-on-xi}
\frac{\log 3}{ \log x_1} = \frac{\log 5}{ \log x_2} = \dots  =\frac{\log (2r+3)}{ \log x_{r+1}}
\text{ with } x_{r+1} := \frac{3}{1- 2 \sum_{i=1}^{r}1/x_{i}}
\end{equation}

One can easily solve these equations numerically by a brute force search for all possible values of $x_1$ in the range between $4$ and $6$ and accuracy $10^{-9}$. This way we get as a solution for $r=3$: 
$$x_1 \approx 4.34480819~~~~x_2 \approx 8.60221014~~~~x_3 \approx 13.48967391~~~~x_4 \approx 18.87735817  $$
For $r=1000$ we obtain:
\begin{equation}
    x_1 \approx 5.92564165~~~~x_2 \approx 13.553044874~~~~\dots ~~~~x_{1001}\approx 222506.653295
    \label{numerical-values}
\end{equation}

To prove a bound on the approximation ratio of the \INRA , we 
will bound the length of the matching 
$\MWPM (V_j)$ computed in line~6 of the algorithm and the total length of all matchings $M_j$
computed in all iterations of the algorithm. We start with a bound on $\ell(\MWPM (V_j))$.

\begin{lemma}
    For input $V \subseteq \R^2$, $\varepsilon > 0$, $r\in \mathbb{N}$ and  $2 < x_1 < x_2 < \ldots <x_r$ 
    that satisfy condition~(\ref{condition-on-xi}) the length of the matching $\MWPM (V_j)$ computed in line~6 of the \INRA\ 
    can be bounded by 
    $$\ell(MWPM(V_j)) ~\le~ (2r+3) \cdot \left(
  |V|^{1/3 +\varepsilon}
 \right) ^{\frac{\log 3}{\log x_1}} \cdot \ell(\MWPM(V)).$$
    \label{lem:lengthofexactmatching}
\end{lemma}

\begin{proof} 
To avoid confusion we denote by $j_0$ the index $j$ after finishing the while-loop
and we use the variable $j$ to denote one of the possible values of the index while running the while-loop. Thus we want to prove a bound on $\ell(MWPM(V_{j_0}))$.

As long as the set $V_j$ has cardinality larger than $|V|^{2/3 -\varepsilon} $
the \INRA\ makes in line~\ref{line:INRA-calls-NRA} a call to the \NRA.
In each call the \NRA\ will execute between~0 and~$r$ calls to the subroutine \EFT.
For $q=0, \ldots, r$ let $a_q$ be the number of calls of the \NRA\ that make exactly $q$ calls to the subroutine \EFT. 
Lemma~\ref{lem:NRA}$(c)$ implies that if the \NRA\ executes $q$ calls to the subroutine \EFT\ with input $V_{j}$, then
\begin{equation}
\ell(\MWPM(V_{j+1})) \le (2q + 3) \cdot \ell(\MWPM(V_{j})). 
\label{eq:MWPM(V_i)-bound}
\end{equation}

Inequality~(\ref{eq:MWPM(V_i)-bound}) yields the following bound
for the length of 
$\MWPM(V_{j_0})$ computed in line~\ref{line:INRA:matching-computation} of the \INRA:
\begin{equation}\label{l(MWPM(V_i)-bound}
\ell(MWPM(V_{j_0})) ~\le~ \prod_{q=0}^r (2q+3)^{a_q} \cdot \ell(\MWPM(V)).
\end{equation}

It remains to get a bound on $\prod_{q=0}^r (2q+3)^{a_q}$.
If the \NRA\ makes $q < r$ calls to the subroutine \EFT\ then from Lemma~\ref{lem:NRA}(a) we know that 
\begin{equation}
    |V_{j+1}| \le |V_j|/x_{q+1}. 
    \label{case q<r}
\end{equation}

If the \NRA\ makes $r$ calls to the subroutine \EFT\ then from Lemma~\ref{lem:NRA}(a) and the definition of $x_{r+1}$ in (\ref{condition-on-xi}) we know that

\begin{equation}
|V_{j+1}| \le |V_j|/ x_{r+1}. 
    \label{case q=r}
\end{equation}

By the condition of the while-loop in lines~\ref{line:INRA:begin-while}--\ref{line:INRA:end-while} 
of the \INRA\ we get by using inequalities~(\ref{case q<r}) and~(\ref{case q=r}) that for the set $V_{j_0}$ in line~\ref{line:INRA:matching-computation} of the \INRA\ we have
 $$
 |V_{j_0}| \le |V|^{2/3-\varepsilon}
 \text{~~ and ~~}
 |V_{j_0}| \le \frac {|V|} {\displaystyle \prod_{q=0}^{r} \left(x_{q+1}\right)^{a_q}}
 $$
Similarly we get for the previous iteration $j_0-1$ as the while-loop has not yet ended and by
inequalities~(\ref{case q<r}) and~(\ref{case q=r}) and the fact that $x_{r+1}$ is the largest 
among all $x_{q+1}$ for $q\in\{0, \ldots, r\}$:
 $$
 |V_{j_0-1}| > |V|^{2/3-\varepsilon}
 \text{~~ and ~~}
 |V_{j_0-1}| \le \frac {|V|\cdot x_{r+1}} {\displaystyle \prod_{q=0}^{r} \left(x_{q+1}\right)^{a_q}}
 $$
which gives
\begin{equation}
|V|^{2/3-\varepsilon} < \frac{|V|\cdot x_{r+1}}{\displaystyle \prod_{q=0}^{r} \left(x_{q+1}\right)^{a_q}} \text{~~ or equivalently ~~}
\displaystyle \prod_{q=0}^{r} \left(x_{q+1}\right)^{a_q} < |V|^{1/3 +\varepsilon} \cdot x_{r+1}.
\label{eq:bound on a_q}
\end{equation}

We now get by using~(\ref{condition-on-xi}) and~(\ref{eq:bound on a_q}):
\begin{eqnarray}
\prod_{q=0}^r (2q+3)^{a_q} 
&  = & \prod_{q=0}^{r} \left(x_{q+1}\right)^{\frac{\log\left(2q+3\right)}{\log x_{q+1}} a_q} \nonumber\\
&   = &  \prod_{q=0}^{r} \left(\left(x_{q+1}\right)^{a_q}\right)^{\frac{\log 3}{\log x_1}}\nonumber\\
 & < & (x_{r+1})^{\frac{\log 3}{\log x_1}} \left(
  |V|^{1/3 +\varepsilon}
 \right) ^{\frac{\log 3}{\log x_1}} \nonumber \\
 &   = &(x_{r+1})^{\frac{\log (2r+3))}{\log x_{r+1}}} \left(
  |V|^{1/3 +\varepsilon}
 \right) ^{\frac{\log 3}{\log x_1}}  \nonumber \\
 &   = &(2r+3) \left(
  |V|^{1/3 +\varepsilon}
 \right) ^{\frac{\log 3}{\log x_1}}
 \label{eq:product-bound}
\end{eqnarray}
By plugging this into inequality~(\ref{l(MWPM(V_i)-bound}) we get:
$$\ell(MWPM(V_j)) ~<~ (2r+3) \cdot \left(
  |V|^{1/3 +\varepsilon}
 \right) ^{\frac{\log 3}{\log x_1}} \cdot \ell(\MWPM(V)).$$
\end{proof}

\begin{lemma}
    For input $V$ and $\varepsilon$, let $t$ denote the number of iterations made by the \INRA.
    Then we have 
    $$\sum_{j=1}^t \ell(M_j) ~\le~ 2\cdot z_t \cdot \prod_{j=1}^{t-1} y_j \cdot \ell(\MWPM(V))$$ where $M_j$ is
    the matching computed in line~3 of the algorithm and $z_j$ and $y_j$  are defined as follows: if 
    the \NRA\ makes $q$ calls to \EFT\ in iteration $j$ of the \INRA\ then 
    $y_j=2q+3$ and $z_j=2q+2$. 
    \label{lem:sumofMi}
\end{lemma}

\begin{proof}
By Lemma~\ref{lem:NRA}(b) we have 
$\ell(M_j) \le z_j \cdot \ell(\MWPM(V_j))$ for all $j=1, \ldots, t$.
Moreover, by Lemma~\ref{lem:NRA}(c) we have $\ell(\MWPM(V_j)) \le y_{j-1} \cdot \ell(\MWPM(V_{j-1}))$ and therefore we get for all $j=1, \ldots, t$:
\begin{equation}
\ell(M_j) ~\le~ z_j \cdot \prod_{k=1}^{j-1} y_k \cdot \ell(\MWPM(V)). 
\label{eq:l(M_i)}    
\end{equation}
We now prove the statement of the lemma by induction on $t$. For $t=1$ we have
by Lemma~\ref{lem:NRA}$(b)$: $\ell(M_1) \le z_1 \cdot \ell(\MWPM(V))$. Now let us assume $t>1$ and that the
statement holds for $t-1$. Using~(\ref{eq:l(M_i)}) we get
\begin{eqnarray*}
\sum_{j=1}^t \ell(M_j) & ~=~ & \ell(M_t) + \sum_{j=1}^{t-1} \ell(M_j) \\
                       & \le & \ell(M_t) + 2\cdot z_{t-1} \cdot \prod_{k=1}^{t-2} y_k \cdot \ell(\MWPM(V)) \\
                       & \le & z_t \cdot \prod_{k=1}^{t-1} y_k \cdot \ell(\MWPM(V)) + 2\cdot z_{t-1} \cdot \prod_{k=1}^{t-2} y_k \cdot \ell(\MWPM(V))\\
                       & = & (z_t \cdot y_{t-1}  + 2 \cdot z_{t-1}) \cdot \prod_{k=1}^{t-2} y_k \cdot \ell(\MWPM(V)).\\ 
\end{eqnarray*}
Now we have $z_t \cdot y_{t-1}  + 2 \cdot z_{t-1} \le z_t \cdot y_{t-1}  + 2 \cdot y_{t-1} \le 2 \cdot z_t \cdot y_{t-1} $
which proves the lemma.
\end{proof}

Combining Lemmas~\ref{lem:lengthofexactmatching} and~\ref{lem:sumofMi}, we can now
state the approximation ratio for the \INRA.

\begin{lemma}
    For a set $V\subseteq \mathbb{R}^2$ of even cardinality,  $\varepsilon > 0$, fixed $r\in \mathbb{N}$, and  $2 < x_1 < x_2 < \ldots < x_r$ 
    that satisfy condition~(\ref{condition-on-xi}) the \INRA\ has approximation ratio 
    $O\left( |V|^{(1/3 +\varepsilon) \cdot \frac{\log 3}{\log x_1} }\right)$.
    \label{lem:approx-ratio-INRA}
\end{lemma}

\begin{proof}
    Similar to the proof of Lemma~\ref{lem:lengthofexactmatching}, 
    let $a_q$ denote the number of iterations in the \INRA\ in which the \NRA\ makes exactly $q$
    calls to the subroutine \EFT. We denote the number of iterations of the \INRA\ by $t$
    and use the definition of $z_j$ and $y_j$ from Lemma~\ref{lem:sumofMi}.
    By Lemma~\ref{lem:sumofMi} we have 
    $$\sum_{j=1}^t \ell(M_j) ~\le~ 2\cdot z_t \cdot \prod_{j=1}^{t-1} y_j \cdot \ell(\MWPM(V)) 
    ~\le~ (4r+4)\cdot \prod_{j=0}^{r} (2q+3)^{a_q}  \cdot \ell(\MWPM(V)).$$ 
    
We already know from inequality~(\ref{eq:product-bound}) in the proof of Lemma~\ref{lem:lengthofexactmatching} that this results in the upper bound $\sum_{j=1}^t \ell(M_j) = O\left(|V|^{(1/3 +\varepsilon) \cdot \frac{\log 3}{\log x_1} }\cdot \ell(\MWPM(V))\right)$. Together with Lemma~\ref{lem:lengthofexactmatching} this implies
that the length of the perfect matching returned by the \INRA\ is bounded by
$O(|V|^{(1/3 +\varepsilon) \cdot \frac{\log 3}{\log x_1} }\cdot \ell(\MWPM(V))).$
 \end{proof}

We can now prove our main result on approximating Euclidean minimum weight perfect matchings for point sets in $\R^2$.

\approxtwodimensional*

\begin{proof}
We claim that our algorithm \INRA\ has the desired properties
if we set $r$ to~1000 and choose  $2 < x_1 < x_2<\ldots < x_{1000}$ that satisfy condition~(\ref{condition-on-xi}). 
From Lemma~\ref{runtime-INRA}
we know that the runtime of this algorithm is $O(n \log n)$. By Lemma~\ref{lem:approx-ratio-INRA}
its approximation ratio is $O\left( |V|^{(1/3 +\varepsilon) \cdot \frac{\log 3}{\log x_1} }\right)$.
Using the numeric solution~(\ref{numerical-values}) for $x_1$ we get
$\frac{1}{3} \frac{\log 3}{\log x_1} < 0.20582$. By choosing $\varepsilon $ sufficiently small we get an approximation ratio of $O(n^{\Rtwoexp})$.
\end{proof}

\section{Extension to Higher Dimensions}
\label{sec:higherDimensions}

We now want to extend our result for the 2-dimensional case to higher dimensions. 
For a fixed dimension $d > 2$ we can use essentially the same approach as in two dimensions,
but need to adjust two things. First, for $d > 2$ no $O(n\log n)$ algorithm is known that 
computes a Euclidean minimum spanning tree in $\R^d$. Instead we will use 
Vaidya's~\cite{Vai1989} $O(n \log n)$ algorithm to compute a $3^{r}$-nearest neighbor graph in $\R^d$ for fixed $d$ and fixed $r$. 
We call this algorithm \EFNN~(Algorithm~\ref{alg:Add-Edges-From-NN-Algorithm}).

\begin{algorithm}[H]
\caption{~~\EFNN}
\label{alg:Add-Edges-From-NN-Algorithm}
\begin{algorithmic}[1]
\Statex \textbf{Input: } a graph $G$ on vertex set $V$, $r\in\mathbb{N}$ and a $3^{r}$-nearest neighbor graph $\NN'$\ on $V$
\Statex \textbf{Output: } a subset of the edges of $\NN'$
\STATE $S := \emptyset$
\STATE Choose a bijection $f:E(\NN')\to \{1,\ldots, |E(\NN')|\}$ s.t.\ $f(e_1) < f(e_2)$ implies $l(e_1) \le l(e_2)$
\FOR{each odd connected component of $G$ with size less than $3^{r}+1$}
     \STATE add to $S$ an edge from $\NN'$  with minimum $f$-value that leaves this component
\ENDFOR
\STATE return the set $S$
\end{algorithmic}
\end{algorithm}

We have to do the following changes to the \NRA~(Algo\-rithm~\ref{alg:Node-Reduction-Algorithm}):
In line~\ref{line:MST-computation}, we calculate a $3^{r}$-nearest-neighbor graph $\NN'$ of $V$ instead of a minimum spanning tree $T$ of $V$. In line~\ref{line:last-line-while-loop}, we call \EFNN \, instead of \EFT. For this modified algorithm, which we call the modified \NRA, we get the following statement, which closely mimics Lemma \ref{lemma:bounds-for-remaining-vertices}.

\begin{lemma}
For a point set $V\subseteq \R^d$, $r \in\mathbb{N}$, and  $x_1$, $x_2$, $\ldots$, $x_r$ satisfying condition~(\ref{condition-on-xi}), let $W$ be the point set and $M$ be the matching returned by the modified \NRA. 
If the subroutine \EFNN\ is called $q$ times within the modified \NRA\ then we have:
\begin{enumerate}
\item  [(a)] $|W|\le $ 
$\begin{cases}
    |V|/x_{q+1} & \text{if } q<r\\
    |V|/ \frac{3}{1 - 2 \sum_{i=1}^{r} \frac{1}{x_i}}   & \text{if } q =r
\end{cases}$
\item  [(b)] $\ell(M)\le (2 q+2) \cdot\ell(\MWPM(V))$
\item  [(c)] $\ell(\MWPM(W))\le (2 q+3)\cdot\ell(\MWPM(V))$
\end{enumerate}
\label{lem:NRA-high-dim}
\end{lemma}
\begin{proof}
    We first prove statement (a). For $q<r$ the proof is the same as in Lemma \ref{lemma:bounds-for-remaining-vertices}. \\
    Case $q=r$.
    While in Section \ref{Sec:node-reduction-algorithm}  we assumed the $r$-well structured sets to be in $\mathbb{R}^2$, the definition clearly extends to higher dimensions if instead of using \EFT\ we use \EFNN.
    The method used in the proof of Lemma~\ref{lemma:structure-of-V} for creating well structured sets does not depend on the dimension but only depends on which components get joined together at which round. 
    All possible combinations can be translated to a line and thus we get a modified version of Lemma~\ref{lemma:structure-of-V} with $V\subseteq \R^d$ and $V' \subseteq \R^d$. 
    Thus Corollary~\ref{cor:well-structured} also holds in higher dimension and we may assume that the point set $V$ is $r$-well structured.
    
    This implies that each odd connected component in each graph $G_i$ up to $i=r-1$ contains at most $3^{r}$ points. 
    Thus, the $3^{r}$-nearest neighbor graph contains for each odd connected component at least one 
    edge leaving it. On $r$-well structured sets the original and the modified \NRA\ therefore produce the same result on $V$. 
    Then by using the Lemma~\ref{lemma:bounds-for-remaining-vertices} for the original \NRA\ we get that $|W|<|V|/\frac{3}{1 - 2 \sum_{i=1}^{r} \frac{1}{x_i}} $.

Statements (b) and (c) follow from the fact that the proof of Lemma \ref{lemma:subroutine-Edges-From-Tree} also holds for the set $S$ returned by \EFNN, meaning that $\ell(S)\le 2\ell(MWPM(V))$. 
This is the only bound that is used for set $S$ in Lemma \ref{lemma:bounds-for-remaining-vertices} for proving the statements (b) and (c), thus the statement also holds for this modified version.

\end{proof}

We also have to change the \INRA~(Algorithm~\ref{alg:Iterated-Node-Reduction-Algorithm}). In line~\ref{line:INRA:begin-while}, instead of the threshold $V^{2/3-\varepsilon}$ we use the threshold $V^{1/3-\varepsilon}$. This is because the fastest known exact algorithm to compute a Euclidean minimum weight perfect matching in $\R^d$ is the $O(n^3)$ implementation of Edmonds' algorithm due to 
Gabow~\cite{Gab1972} and Lawler~\cite{lawler2001combinatorial}. Using these two changes we get:

\approxhigherdimension*
\begin{proof}
We get a higher dimensional analog of Lemma~\ref{lem:lengthofexactmatching} 
by plugging Lemma~\ref{lem:NRA-high-dim} instead of Lemma~\ref{lem:NRA} into the proof of Lemma~\ref{lem:lengthofexactmatching}.
In addition
we replace $|V|^{1/3+\varepsilon}$ by $|V|^{2/3+\varepsilon}$. If we apply all these changes then the approximation ratio in Lemma~\ref{lem:approx-ratio-INRA} changes to $O\left(|V|^{\frac23 \frac{\log 3}{\log x_1}}\right)$.
Now setting $r=1000$ and choosing the numeric value~(\ref{numerical-values}) for $x_1$ we get
$\frac23 \frac{\log 3}{\log x_1} < 0.41163$. This implies the claimed approximation ratio $O(n^{\Rdexp})$. The runtime analysis is the same as that from Lemma~\ref{runtime-INRA} 
due to Vaidya's $O(n\log n)$ algorithm~\cite{Vai1989} for computing \NN'.
\end{proof}

Vaidya~\cite{Vai1989} states the runtime of his nearest-neighbor algorithm as 
$O((cd)^dn\log n )$, making the dependence on the dimension $d$ explicit.
Since computing \NN' is the most time-consuming step in our approach, the overall runtime of our algorithm has exactly the same dependence on 
$d$ as Vaidya’s algorithm.

\textbf{Remark.} Theorem~\ref{thm:approxhigherdimension}  also holds if instead of the Euclidean metric, we have some other $L_p$, $p=1,2,\ldots, \infty$ metric. 
The runtime stays the same, as we can do all the computational steps, including the computationally expensive steps like finding the $3^{r}$-nearest neighbor graph with the same runtime according to~\cite{Vai1989}. 
The approximation ratio also stays the same, as when finding different bounds, we only use the triangle inequality which also holds in $L_p$ metric spaces.

\section{A Lower Bound Example}
\label{sec:lowerbound}

In this section we provide a lower bound example for the \INRA , showing that its approximation 
ratio cannot be better than $O(n^{0.106})$.

\begin{lemma}
    The approximation ratio of the \INRA\ cannot be better than $\Omega(n^{0.106})$.
\end{lemma}

\begin{proof}

We recursively define point sets $V_i$ for $i=0,1,2,\ldots$ as follows. 
All points in $V_i$ lie on a horizontal line. The set $V_0$ contains two points at distance $1$.
For $i\ge 1$ the point set $V_i$ is obtained by placing seven copies of $V_{i-1}$ next to each other on a horizontal 
line at distance $13^{i-1}$. Figure~\ref{fig:point-sets-Vi} shows the point sets $V_0$, $V_1$, and $V_2$.

\begin{figure}[ht]
\begin{tikzpicture}[scale=0.083]
    % We use a counter 'x' to keep track of the current x-position.
    \newcounter{x}
    \setcounter{x}{0}
    
    \draw[very thin, gray] (0,15) -- (1,15);
    \fill (0, 15) circle (7pt);
    \fill (1, 15) circle (7pt);

   \draw[very thin, gray] (0,7.5) -- (13,7.5);
    \foreach \i in {1,...,14} {
        \fill (\value{x}, 7.5) circle (7pt);
        \addtocounter{x}{1}
   }

    \setcounter{x}{0}

   \draw[very thin, gray] (0,0) -- (169,0);
    \foreach \i in {1,...,98} {
        \fill (\value{x}, 0) circle (7pt);
        
        % For all but the last point, increase 'x' by 9 if i = 10,20,30,40; otherwise increase by 1.
        \ifthenelse{\i<98}{
            \ifthenelse{\i=14 \OR \i=28 \OR \i=42 \OR \i=56 \OR \i=70 \OR \i=84}
                       {\addtocounter{x}{13}}
                       {\addtocounter{x}{1}}
        }{}
    }
   \node[] (dummy) at (-4.5, 0) {$V_2$};
   \node[] (dummy) at (-4.5, 7.5) {$V_1$};
   \node[] (dummy) at (-4.5, 15) {$V_0$};
\end{tikzpicture}
\caption{The recursively constructed point sets $V_i$ shown for $i=0,1,2$.}
\label{fig:point-sets-Vi}
\end{figure}

Immediately from the recursive definition of the sets $V_i$ we get $|V_i| = 2\cdot 7^i$.
Each set $V_i$ for $i\ge 1$ consists of $7^{i-1}$ groups of fourteen points placed at larger distance. 
As the distance between two neighboring points within each of these groups of fourteen points is exactly~1,
we find a perfect matching of length~7 within each such group of fourteen points.
Therefore we have $\ell(MWPM(V_i)) = 7^i = |V_i|/2$ for all $i \ge 0$.

\begin{figure}[ht]
\begin{tikzpicture}[scale=0.5]
    % We use a counter 'x' to keep track of the current x-position.
    \setcounter{x}{0}

   %\draw[thin, gray] (0,5) -- (9,5);
   \draw[thick,red] (0,5) -- (6,5) (7,5)--(13,5);
    \foreach \i in {1,...,14} {
        \ifthenelse{\i=1 \OR \i=14}
        {\fill[red] (\value{x}, 5) circle (5pt);}
        {\fill (\value{x}, 5) circle (5pt);}
        \addtocounter{x}{1}
   }
  
   \node[] (dummy) at (-3, 5) {$V_1$};
\end{tikzpicture}
\caption{A possible nearest neighbor graph for the point set $V_1$. The edges of the nearest neighbor graph are shown in red. The two points shown in red are the points that the \NRA\ may return as the set $W$.}
\label{fig:NN(V1)}
\end{figure}

If the \INRA\ is started on input set $V_i$ for $i\ge 1$ then $\NN(V_i)$ may contain exactly twelve edges within each group of fourteen consecutive points, as shown for $V_1$ in Figure~\ref{fig:NN(V1)}. 
Then the graph $G_0$ in the \NRA\ will contain $2\cdot 7^{i-1} = |V_i|/7$ connected components each of size~7. 
If we choose $x_1 \approx 5.9256$
according to~(\ref{numerical-values}) then as $G_0$ has $|V_i|/7 < |V_i| / x_1$ many components,
the routine \EFT\ is not called.
Now the \NRA\ may choose within each group of fourteen consecutive points the two outermost points 
to include in the set $W$ (see the example in Figure~\ref{fig:NN(V1)}).
Thus, if the \NRA\ is started on the set $V_i$, then the set $W$ after one iteration is 
a scaled version of the set $V_{i-1}$ where the scaling factor is~$13$. 
Thus, the total length of a matching returned by the \INRA\ on the point set $V_i$ is $ \frac27 |V_i|$
plus the length of a matching returned by the \INRA\ on the point set $V_{i-1}$ scaled by a factor of~$13$.
The \INRA\ stops as soon as the remaining set of points has size less than $|V_i|^{2/3-\varepsilon}$.
As in each iteration the size of the remaining point set decreases by a factor of~$7$, we will have
at least $k$ iterations where $k$ satisfies $|V_i|/7^k = |V_i|^{2/3}$. This results in 
$ k = \frac13\log_7|V_i|$. Therefore, we get as a lower bound for the length of the matching returned by the \INRA\ on the point set $V_i$:

$$ \frac27 \cdot |V_i| + \frac27 \cdot 13 \cdot \frac{|V_i|}{7} + \ldots + \frac27 \cdot 13^{\lfloor k\rfloor }\cdot  \frac{|V_i|}{7^{\lfloor k\rfloor}} = 
\frac27\cdot |V_i| \cdot\sum_{j=0}^{\lfloor k\rfloor} \left(\frac{13}{7}\right)^j = 
\frac13 |V_i| \left(\left(\frac{13}{7}\right)^{\lfloor k\rfloor+1}-1\right).$$
Using $k = \frac13\log_7|V_i|$ and assuming $k$ to be sufficiently large we get 

$$\frac{|V_i|}{3} \left(\left(\frac{13}{7}\right)^{\lfloor k\rfloor+1}-1\right) \ge 
\frac{|V_i|}{4} \left(\frac{13}{7}\right)^{ \frac13\log_7|V_i|} = 
\frac{|V_i|}{4} |V_i|^{ \frac13 \frac{\log (13/7)}{\log 7}} > \frac{|V_i|^{1.106}}{4} 
$$

Thus we get an approximation ratio of $\Omega(n^{0.106})$.
The total distance between the two outer most points in $V_i$ is at most $|V_i|^{1.32}$.
So the set $V_i$ and its distances can be encoded with polynomial size. 

\end{proof}

\section*{Note Added in Proof}  The first deterministic $n/2$ approximation algorithm for the Euclidean minimum weight perfect matching problem with runtime $O(n \log n)$  was presented by Papadimitriou. It is described in~\cite{SPR1980}. The authors state that the runtime of the algorithm is $O(n^2)$ in the metric case. By applying the result of~\cite{SH1975} one gets $O(n\log n)$ runtime in the Euclidean case. Reingold and Tarjan~\cite{RT1981} showed that the greedy heuristic for the Euclidean minimum weight perfect matching problem has approximation ratio $n^{0.58496}$. They state a runtime of $O(n^2 \log n)$. By applying the result of Bespamyatnikh~\cite{Bes1998} one gets  $O(n \log n)$ runtime.

%\section*{Declarations}
%\paragraph{Competing Interests} The authors have no relevant financial or non-financial interests to disclose.

\bibliographystyle{plainurl}
\bibliography{ref}

\end{document}